\documentclass[aps,pra,reprint,amsmath,amssymb,amsfonts,showkeys,showpacs,preprintnumbers]{revtex4-1}
\usepackage{bm,graphicx,xcolor,hyperref,rotating,lineno,multirow}
\newcommand{\mc}{\multicolumn}
\newcommand{\mr}{\multirow}
\newcommand{\Ec}{E_{\rm c}}
\newcommand{\mEh}{{\rm m}E_{\rm h}}

\begin{document}

\title{Correlation energy of anisotropic quantum dots}

\author{Yan Zhao}
\email{yan.zhao@anu.edu.au}
\author{Pierre-Fran\c{c}ois Loos}
\email{loos@rsc.anu.edu.au}
\author{Peter M. W. Gill}
\thanks{Corresponding author}
\email{peter.gill@anu.edu.au}
\affiliation{
Research School of Chemistry, 
Australian National University, 
Canberra, ACT 0200, Australia}
\date{\today}

\begin{abstract}
We study the $D$-dimensional high-density correlation energy $\Ec$ of the singlet ground state of two electrons confined by a harmonic potential with Coulombic repulsion.  We allow the harmonic potential to be anisotropic, and examine the behavior of $\Ec$ as a function of the anisotropy $\alpha^{-1}$.  In particular, we are interested in the limit where the anisotropy goes to infinity ($\alpha\to0$) and the electrons are restricted to a lower-dimensional space.  We show that tuning the value of $\alpha$ from $0$ to $1$ allows a smooth dimensional interpolation and we demonstrate that the usual model, in which a quantum dot is treated as a two-dimensional system, is inappropriate.  Finally, we provide a simple function which reproduces the behavior of $\Ec$ over the entire range of $\alpha$.
\end{abstract}


\maketitle

\section{
\label{sec:intro}
Introduction}

The two-electron problem is one of the fundamental problems of quantum physics \cite{BetheSalpeter, Hylleraas29, Hylleraas30, Hylleraas64} and, although it looks simple, it has only been solved in certain very special cases \cite{Santos68, Moshinsky68, Kais89, Alok90, Kais93, Taut93, LoosHook, Quasi09, LoosExcitSph}.  Many of the methods that have been developed to provide approximate solutions to the two-electron problem have been central in the development of molecular physics and quantum chemistry \cite{Kohn99, Pople99}.

The familiar Hartree-Fock (HF) model \cite{Szabo} treats a system as a separable collection of electrons, each moving in the mean field of the others.  The HF solution provides us with a good approximation to the energy and is widely applied to model complex molecular systems \cite{Helgaker}.  However, it is essential to understand its error \cite{Lowdin59}
\begin{equation}
	\Ec = E - E_{\text{HF}},
\end{equation}
which Wigner called the correlation energy \cite{Wigner34}.  Studies of correlation effects in two-electron systems are interesting in their own right, but also provide simple examples to test computational models \cite{ParrYang} and shed light on more complicated systems \cite{ONeill03, TEOAS09, LoosConcentric}.  They have been extensively studied, for various confining external potentials, interacting potentials and degrees of freedom \cite{EcLimit09, LoosBall10, EcProof10, Frontiers}.

However, most previous studies have focussed on spherically symmetric external potentials, for anisotropy significantly complicates the mathematical analysis.  This is unfortunate, for most real systems are not isotropic, and it is therefore important to understand how anisotropy affects the correlation energy.

\section{
\label{sec:QD}
Quantum dots at high density}

Quantum dots are often modeled by electrons in a harmonic potential with Coulombic repulsion \cite{Kestner62, Alhassid00, Ando82, Reimann02}.  Because experimental conditions strongly confine the electrons in one dimension, the model potentials are usually spherical and two-dimensional.  Calculations on such quantum dots have been used extensively in the development of exchange-correlation density functionals for low-dimensional systems in the framework of density-functional theory (DFT) \cite{Pittalis07, Helbig08, Pittalis09, Pittalis10, Rasanen10, Sikaroglu10}.

In addition to experimental progress \cite{Ashoori93, Tarucha96, Kouwenhoven97}, many theoretical investigations have studied the effects of the confinement strength in the third dimension on the energy of the quantum dot.  These studies have used DFT \cite{Lee98, Jiang01}, HF \cite{Fujito96}, exact diagonalization \cite{Sun03} and exact solutions \cite{Lin01, Zhu05, Liu07}.  However, despite the importance of the correlation energy, only a few studies \cite{Sloggett05, Dong05, Waltersson07} have explored the confinement effect on $\Ec$.

In this paper, we examine the effects of anisotropy on the energy of the nodeless ground state of two electrons in a $D$-dimensional harmonic potential, using atomic units throughout.  We define the external potential by
\begin{equation} 
	V(\bm{r}) = \frac{\lambda^4}{2} \sum_{j=1}^D \frac{x_j^2}{\alpha_j^4},
\end{equation}
where $\lambda$ governs the overall strength of the potential and $\alpha_j \in (0,1]$ is the force constant for the Cartesian coordinate $x_j$.  The isotropic case is obtained when all $\alpha_j$'s are equal.  We are particularly interested in the behavior where one or more of the $\alpha_j$ approach $0$ for, in such limits, the system is constrained towards a lower dimensionality.

We restrict our attention to the high-density ($\lambda\to\infty$) limit \cite{GellMann57, White70, Benson70, Vignale} for it has been found that the high-density behavior of electrons is surprisingly similar to that at typical atomic and molecular electron densities \cite{EcLimit09, LoosBall10, EcProof10, Frontiers}.

The Hamiltonian describing this system is
\begin{equation}
	\Hat{H} = - \frac{1}{2} \left(\nabla_1^2 + \nabla_2^2\right) + V(\bm{r}_1) + V(\bm{r}_2) + \frac{1}{r_{12}},
\end{equation}
where $r_{12}=|\bm{r}_1-\bm{r}_2|$ is the interelectronic distance and $\nabla^2$ is the $D$-dimensional Laplace operator \cite{Herschbach86, Avery91}.  Scaling all lengths by $\lambda$, we obtain
\begin{equation} \label{generalH}
	\Hat{H} = \lambda^2 \left[-\frac{\nabla_1^2}{2} - \frac{\nabla_2^2}{2} + V(\bm{r}_1) + V(\bm{r}_2) + \frac{1}{\lambda}\frac{1}{r_{12}}\right],
\end{equation}
which is suitable for large-$\lambda$ perturbation theory with the zeroth-order Hamiltonian 
\begin{equation}
	\Hat{H}^{(0)} = - \frac{\nabla_1^2}{2} - \frac{\nabla_2^2}{2} + V(\bm{r}_1) + V(\bm{r}_2),
\end{equation}
and the perturbation
\begin{equation}
	\Hat{H}^{(1)} = \frac{1}{r_{12}}.
\end{equation}

The Hamiltonian $\Hat{H}^{(0)}$ is separable and its eigenfunctions and energies are
\begin{multline}
	\Psi_{\bm{n}_1,\bm{n}_2}^{(0)}(\bm{r}_1,\bm{r}_2) = \frac{1}{\pi^{D/2}}  \prod_{j=1}^D \frac{\exp[-(x_{1,j}^2+x_{2,j}^2)/(2\alpha_j^2)]}
																						{2^{(n_{1,j}+n_{2,j})/2} \alpha_j \sqrt{n_{1,j}! n_{2j}!}}
	\\
	\times H_{n_{1,j}}(x_{1,j}/\alpha_j) H_{n_{2,j}}(x_{2,j}/\alpha_j),
\end{multline}
and
\begin{equation}
	E_{\bm{n}_1,\bm{n}_2}^{(0)} 
	= \sum_{j=1}^D \frac{n_{1,j}+n_{2,j}+1}{\alpha_j^2},
\end{equation}
where $\bm{n}_i=(n_{i,1},\ldots,n_{i,D})$ holds the quantum numbers of the $i$th electron, $x_{i,j}$ is the $i$th coordinate of the $j$th electron, and $H_n(x)$ is the $n$th Hermite polynomial \cite{NISTbook}.

Expanding the exact and HF energies as power series
\begin{align}
	E		& = \lambda^2 E^{(0)} + \lambda E^{(1)} 
			+ E^{(2)} + O(\lambda^{-1}),\\
	E_{\text{HF}}	& = \lambda^2 E_{\text{HF}}^{(0)} 
			+ \lambda E_{\text{HF}}^{(1)} 
			+ E_{\text{HF}}^{(2)} + O(\lambda^{-1}),
\end{align}
one finds \cite{White70, Benson70, Katriel05, HookCorr05} that
\begin{align}
	E^{(0)} & = E_{\text{HF}}^{(0)},&
	E^{(1)} & = E_{\text{HF}}^{(1)},
\end{align}
and, therefore, that the limiting correlation energy is
\begin{equation}
\label{Ec-def}
	\Ec = \lim_{\lambda\to\infty} \left(E - E_{\text{HF}}\right) = E^{(2)} - E_{\text{HF}}^{(2)}.
\end{equation}

In this paper, we show that $\Ec$ is strongly affected by the anisotropy and dimensionality of the potential.  In Sec.~\ref{sec:AH}, we use perturbation theory to obtain integral expressions for $E^{(2)}$ and $E_{\text{HF}}^{(2)}$ in an anisotropic quantum dot.  In Sec.~\ref{sec:solutions}, we use the integral  to express $\Ec$ as a infinite sum in a special case.  Finally, in Sec.~\ref{sec:numerical}, we present numerical results and discuss some of the implications with regard to quantum dots and dimensional interpolation.

\section{
\label{sec:AH}
Second-order energies}

The exact and HF second-order energies are \cite{HookCorr05,EcLimit09}
\begin{align}
	E^{(2)}				& = \sum_{(\bm{m},\bm{n})\ne0} \frac{\left<\Psi_{\bm{m},\bm{n}}^{(0)}\left|r_{12}^{-1}\right|\Psi_{0,0}^{(0)}\right>^2}
																{E_{\bm{m},\bm{n}}^{(0)}-E_{0,0}^{(0)}},	\\
	E_{\text{HF}}^{(2)}	& = 2\sum_{\bm{m}\ne0}\frac{\left<\Psi_{\bm{m},0}^{(0)}\left|r_{12}^{-1}\right|\Psi_{0,0}^{(0)}\right>^2}
																{E_{\bm{m},0}^{(0)}-E_{0,0}^{(0)}}.
\end{align}
Whereas the summation for the exact energy includes all states, the summation for the HF energy includes only singly-excited states \cite{HookCorr05}.

Employing the Fourier representation
\begin{equation}
	\frac{1}{r_{12}} = \frac{\Gamma\left(\frac{D-1}{2}\right)} {2\pi^{(D+1)/2}} \int \frac{d\bm{k}}{k^{D-1}} e^{i\bm{k}\cdot(\bm{r}_1-\bm{r}_2)},
\end{equation}
one finds
\begin{align}
\label{E2-3}
	E^{(2)} = & -\frac{\Gamma\left(\frac{D-1}{2}\right)^2} {4\pi^{D+1}} \sum_{\bm{n}\ne\bm{0}} \frac{\prod_{j=1}^D (-\alpha_j^2)^{n_j}/n_j!}
				{\sum_{j=1}^D n_j/\alpha_j^2}	
	\notag
	\\
	& \times \left[\int \frac{d\bm{k}}{k^{D-1}} \prod_{j=1}^D k_j^{n_j} \exp(-k_j^2 \alpha_j^2 / 2)\right]^2,
	\\
	\label{E2HF-3}
	E_{\text{HF}}^{(2)} = & - \frac{\Gamma(\frac{D-1}{2})^2} {2\pi^{D+1}} \sum_{\bm{n}\ne\bm{0}} \frac{\prod_{j=1}^D (-\alpha_j^2/2)^{n_j}/n_j!}
				{\sum_{j=1}^D n_j/\alpha_j^2}
	\notag
	\\
	& \times \left[\int \frac{d\bm{k}}{k^{D-1}} \prod_{j=1}^D k_j^{n_j} \exp(-k_j^2 \alpha_j^2 / 2) \right]^2,
\end{align}
where $\Gamma$ is the gamma function \cite{NISTbook}.

\section{
\label{sec:solutions}
Correlation energy}

We now try to solve the integral 
\begin{equation} \label{integral}
	\mathcal{K} = \int \frac{d\bm{k}}{k^{D-1}} \prod_{j=1}^D k_j^{n_j} \exp(-k_j^2 \alpha_j^2 / 2)
\end{equation}
for special values of $\alpha_j$.  The isotropic case, {\em i.e.}~where all $\alpha_j$ are equal, has been considered in detail in Ref.~\cite{EcLimit09}.  In the present paper, we generalize this to case where the $\alpha_j$ take two distinct values.  Without loss of generality, we let $\delta$ be an integer such that $0<\delta<D$, and set 
\begin{gather}
	\alpha_1 = \cdots = \alpha_\delta = \alpha,\\
	\alpha_{\delta+1} = \cdots= \alpha_D = 1.
\end{gather}
We assume $\alpha\in(0,1]$, {\em i.e.}~the potential is stronger in the first $\delta$ dimensions.

The integral \eqref{integral} vanishes if any of the $n_i$ is odd.  When all are even, it can be evaluated in hyperspherical coordinates \cite{Louck60} and one finds
\begin{multline}
\label{integralhg}
	\mathcal{K} = \frac{2^{\frac{n+1}{2}} \Gamma(\frac{n+1}{2})}{\Gamma(\frac{n+D}{2})}
					\prod_{j=1}^D \Gamma\left(\frac{n_j+1}{2}\right)	\\
	\times F\left[\frac{n+1}{2} , \frac{m+\delta}{2} , \frac{n+D}{2} ; 1-\alpha^2\right],
\end{multline}
where $F$ is the Gauss hypergeometric function \cite{NISTbook} and
\begin{align}
	m & = \sum_{j=1}^\delta n_j,	& n & = \sum_{j=1}^D n_j.
\end{align}

Since we only need to sum over terms where $n_j$ is even for all $1\le j\le D$, we can replace $n_j\to 2n_j$ throughout.  Substituting Eq.~\eqref{integralhg} into Eqs.~\eqref{E2-3} and \eqref{E2HF-3} yields
\begin{equation}
\begin{split}
\label{Ec-4}
	\Ec(\alpha,D,\delta)
	& = \frac{\Gamma\left(\frac{D-1}{2}\right)^2}
	{4\pi\Gamma\left(\frac{\delta}{2}\right)\Gamma\left(\frac{D-\delta}{2}\right)}
	\\
	& \quad \times \sum_{n=1}^{\infty}
	\left(1-\frac{1}{2^{2n-1}}\right)
	\frac{\Gamma\left(n+\frac{1}{2}\right)^2}
	{\Gamma\left(n+\frac{D}{2}\right)^2}
	\\
	& \quad \times \sum_{m=0}^{n} 
	\frac{\alpha^{4m+2}\Gamma\left(m+\frac{\delta}{2}\right)\Gamma\left(n-m+\frac{D-\delta}{2}\right)}
	{m!(n-m)!\left[\alpha^2n+(1-\alpha^2)m\right]}
	\\
	& \quad \times F \left[n+\frac{1}{2} , m+\frac{\delta}{2} , n+\frac{D}{2} ; 1-\alpha^2\right]^2.
\end{split}
\end{equation}

In the isotropic limit ($\alpha=1$), this becomes
\begin{equation}
\begin{split}
	\Ec(1,D,\delta) & = 
	-\frac{\Gamma\left(\frac{D-1}{2}\right)^2}{4\pi\Gamma\left(\frac{D}{2}\right)}
	\\
	& \quad \times
	\sum_{n=1}^{\infty} 
	\left(1-\frac{1}{2^{2n-1}}\right)
	\frac{\Gamma\left(n+\frac{1}{2}\right)^2}{\Gamma\left(n+\frac{D}{2}\right)n!n}
	\\
	& = \Ec(D),
\end{split}
\end{equation}
as presented in Ref.~\cite{EcLimit09}.

In the anisotropic limit ($\alpha\to0$), we must consider two cases.  If $\delta=D-1$, \eqref{Ec-4} becomes infinite, because the second-order energies and the correlation energy are unbounded for the one-dimensional dot \cite{Doren87, Herschbach86, Goodson87, Morgan93}.  However, if $\delta<D-1$, we have
\begin{equation}
\label{Ec0} 
\begin{split}
	\Ec(0,D,\delta)
	& = -\frac{\Gamma\left(\frac{D-\delta-1}{2}\right)^2}{4\pi\Gamma\left(\frac{D-\delta}{2}\right)}
	\\
	& \quad \times \sum_{n=1}^{\infty}
	\left(1-\frac{1}{2^{2n-1}}\right)
	\frac{\Gamma\left(n+\frac{1}{2}\right)^2}{\Gamma\left(n+\frac{D-\delta}{2}\right)n!n}
	\\
	& = \Ec(D-\delta),
\end{split}
\end{equation}
which confirms that, as the electrons are squeezed from a $D$-dimensional space to a $(D-\delta)$-dimensional space, their correlation energy tends smoothly toward the expected value for $(D-\delta)$ dimensions.

\section{
\label{sec:numerical}
Numerical Results and Discussion}

\begin{table}
\caption{
\label{tab:Ectable}
Limiting correlation energies ($\mEh$) for quantum dots with various $\alpha$, $D$ and $\delta$.}
\begin{ruledtabular}
\begin{tabular}{ccccccccc}
$D$	&	$\delta$	&	\mc{7}{c}{Anisotropy $\alpha$}
\\ 
\cline{3-9}
	&		&	0			&	1/32		&	1/16		&	1/8			&	1/4			&	1/2			&	1		\\
\hline
\mr{1}{*}{3}
	&	1	&	239.6		&	210.3		&	187.4		&	152.8		&	109.3		&	68.6		&	49.7	\\
\\
\mr{2}{*}{4}
	&	1	&	49.7		&	48.9		&	47.4		&	43.8		&	36.6		&	26.4		&	19.9	\\
	&	2	&	239.6		&	195.8		&	163.9		&	119.7		&	70.8		&	33.3		&	19.9	\\
\\
\mr{3}{*}{5}
	&	1	&	19.9		&	19.8		&	19.5		&	18.8		&	16.9		&	13.3		&	10.4	\\
	&	2	&	49.7		&	48.3		&	45.7		&	39.7		&	29.2		&	16.7		&	10.4	\\
	&	3	&	239.6		&	185.8		&	148.7		&	100.3		&	51.8		&	19.9		&	10.4	\\
\\
\mr{4}{*}{6}
	&	1	&	10.4		&	10.4		&	10.4		&	10.1		&	9.4			&	7.9			&	6.4		\\
	&	2	&	19.9		&	19.7		&	19.3		&	17.9		&	14.8		&	9.7			&	6.4		\\
	&	3	&	49.7		&	47.7		&	44.2		&	36.6		&	24.2		&	11.6		&	6.4		\\
	&	4	&	239.6		&	178.0		&	137.5		&	87.0		&	40.4		&	13.2		&	6.4		\\
\\
\mr{5}{*}{7}
	&	1	&	6.4			&	6.4			&	6.3			&	6.3			&	5.9			&	5.1			&	4.3		\\
	&	2	&	10.4		&	10.4		&	10.3		&	9.8			&	8.6			&	6.2			&	4.3		\\
	&	3	&	19.9		&	19.6		&	19.0		&	17.2		&	13.1		&	7.4			&	4.3		\\
	&	4	&	49.7		&	47.1		&	42.9		&	34.0		&	20.7		&	8.6			&	4.3		\\
	&	5	&	239.6		&	171.6		&	128.6		&	77.2		&	32.8		&	9.4			&	4.3		\\
\end{tabular}
\end{ruledtabular}
\end{table}

\begin{figure}
	\includegraphics[width=0.4\textwidth]{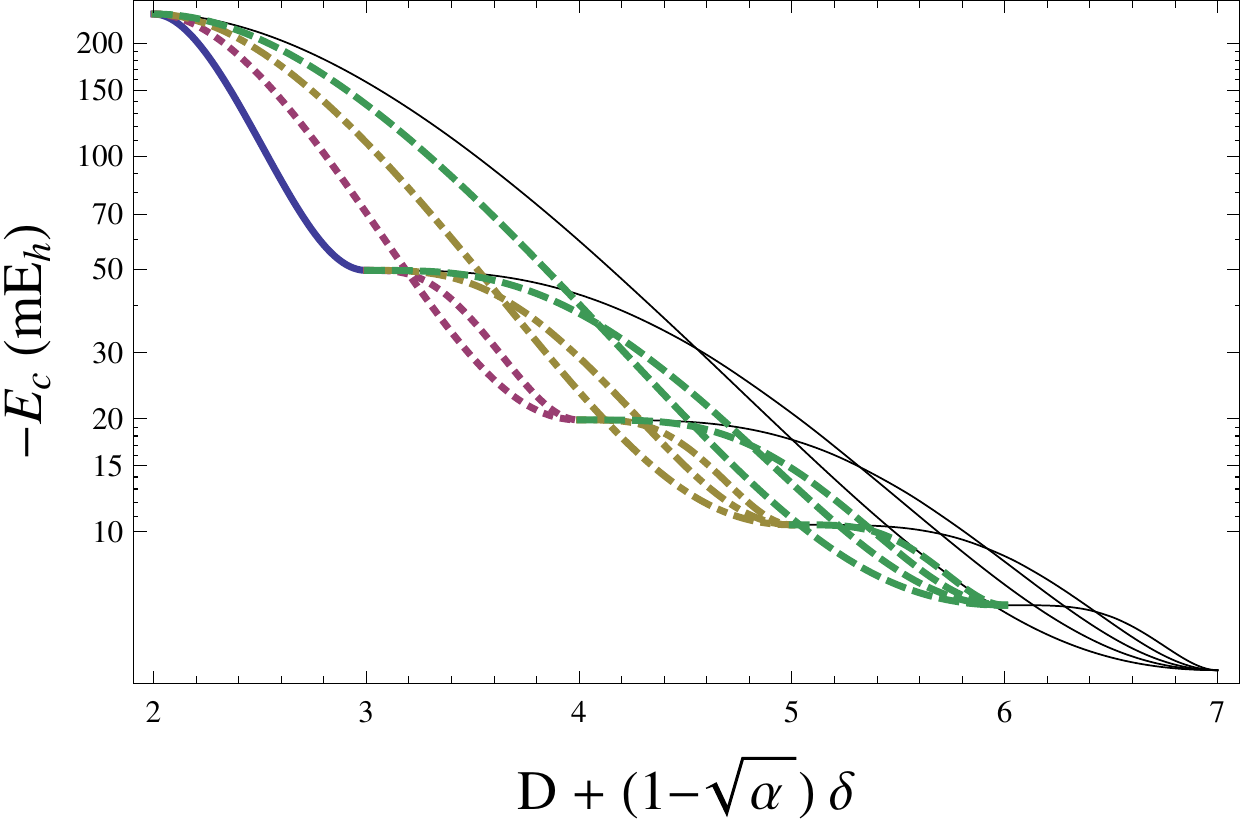}
\caption{
\label{fig:Eplots}
Variation of $-\Ec$ with $D+(1-\sqrt{\alpha})\delta$, showing the correlation ``pathways'' that start from $D=3$ (solid blue), $D=4$ (dotted red), $D=5$ (dot-dashed yellow), $D=6$ (dashed green) and $D=7$ (solid black).}
\end{figure}

The correlation energies of quantum dots with various anisotropies $\alpha$, for $D=3,\ldots,7$ and $\delta=1,\ldots,D-2$ are presented in Table \ref{tab:Ectable}.  The energies for the spherically-symmetric states ($\alpha=0,1$) coincide with the closed-form expressions reported in Ref.~\cite{Frontiers}.  When $\delta$ increases, the variation in correlation energy around $\alpha=1/2$ becomes flatter, but more pronounced near $\alpha=0$.  In contrast, near $\alpha=1$, $\Ec$ remains independent of $\delta$.  Although not unique, pathways between dimensionalities are smooth, and provide proper dimensional interpolations \cite{Loeser85, Herschbach86}.

Figure \ref{fig:Eplots} shows the different correlation ``pathways'' when one starts from $D=3,\ldots,7$ to reach $\delta=1,\ldots,D-2$; the lower and the upper curves correspond to the smaller ($\delta=1$) and larger ($\delta=D-2$) values of $\delta$, respectively.

\begin{figure}
	\includegraphics[width=0.4\textwidth]{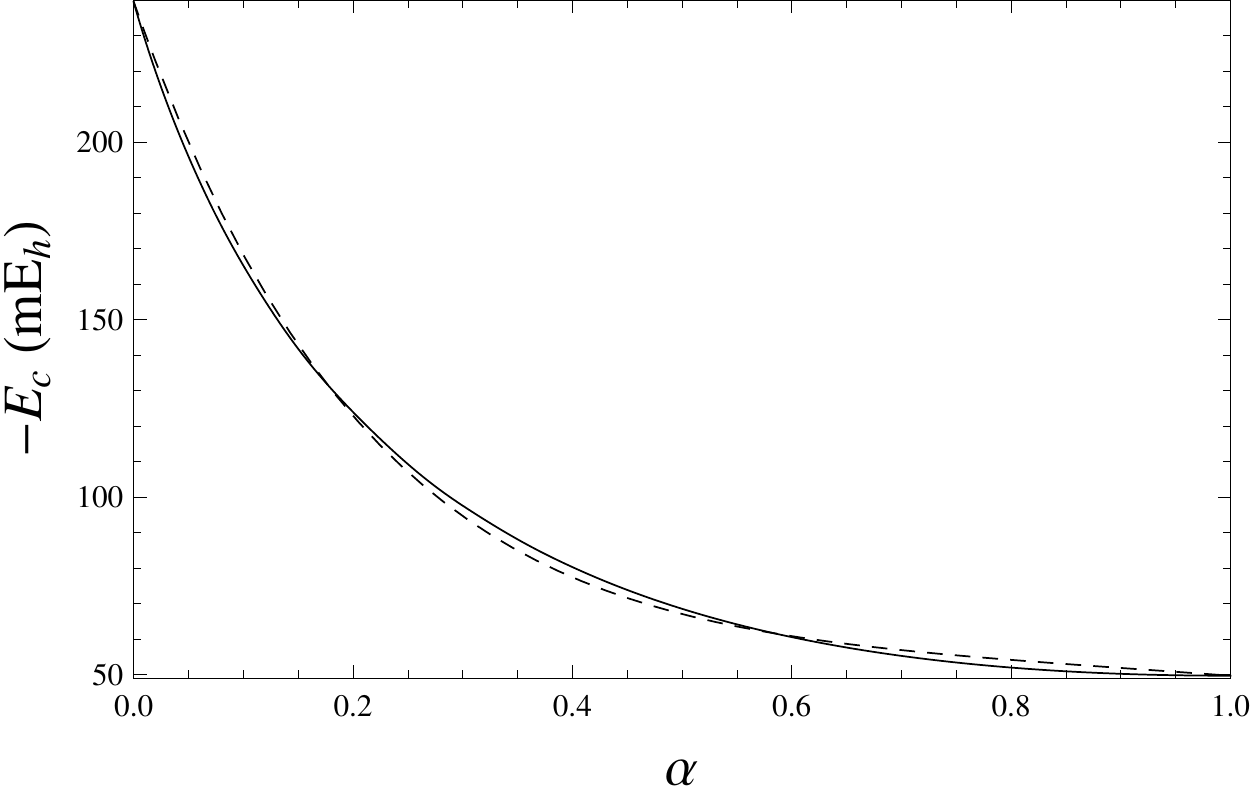}
\caption{
\label{fig:EcD3R1}
Variation of $-\Ec(\alpha,3,1)$ with $\alpha$ (solid).  The fitting equation \eqref{EcApprox} is also shown (dashed).}
\end{figure}

The $D=3$ and $\delta=1$ case models a quasi-2D quantum dot, confined in a three-dimensional harmonic well whose force constant is higher in one of the dimensions.  However, we see from Fig.~\ref{fig:EcD3R1} that, even when the anisotropy is large ($\alpha \ll 1$), the correlation energy is significantly smaller than the $D=2$ limit, which is frequently used to approximate the correlation energy of the quantum dot.  Indeed, for $\alpha>1/2$, the correlation energy is much better approximated by the $D=3$ limit.

To illustrate the difference in correlation energy between a strict and quasi 2D quantum dot, one can calculate $\Ec\left(\alpha,3,1\right)$ for various values of $\alpha$ found in the literature.  In this way, we have found $\Ec = 73.8$ for $\alpha = 0.43$ \cite{Ashoori93, Lin01}, $\Ec = 88.6$ for $\alpha = 0.33$ \cite{Ashoori93, Fujito96} and $\Ec = 110.0$ $\mEh$ for $\alpha = 0.24$ \cite{Tarucha96, Jiang01}.  Even for the smallest of these $\alpha$ values, the correlation energy is less than half of the 2D limit (239.6 $\mEh$).

The correlation energy $\Ec(\alpha,3,1)$ (in $\mEh$) can be accurately approximated using
\begin{equation}
\label{EcApprox}
	\Ec\left(\alpha,3,1\right) \approx 
	\left(c_0 + c_1 \alpha + c_2 \alpha^2\right) e^{-\zeta \alpha},
\end{equation}
with
\begin{align}
	c_0 & = 239.6,&
	c_1 & = -360.3,\nonumber\\
	c_2 & = 557.5,&
	\zeta & = 2.1736,
\end{align}
as shown in Fig.~\ref{fig:EcD3R1}.  By construction, this approximation is exact for $\alpha=0$ and $\alpha=1$.

\section{Conclusion}

In this paper, we have studied the electron correlation of anisotropic quantum dots in the high-density limit.  Using perturbation theory, we have solved the general Hamiltonian \eqref{generalH}, and we have obtained integral expressions for $E^{(2)}$ and $E_{\text{HF}}^{(2)}$.  In the case where $\delta$ dimensions are scaled by a factor of $\alpha$ with respect to the remaining $(D-\delta)$ dimensions, we have expressed the correlation energy $\Ec$ as a infinite sum.

Our numerical results reveal that $\Ec$ remains similar to that of the $D$-dimension system for most $\alpha<1$, only increasing to that of the $(D-\delta)$-dimensional state near $\alpha=0$.  In the physically important $D=3$ and $\delta=1$ case, the correlation energy is well approximated by the $D=2$ limit only if $\alpha \lesssim 0.1$.  Such extreme anisotropy is probably difficult to realize experimentally.

\begin{acknowledgments}
P.M.W.G. thanks the NCI National Facility for a generous grant of supercomputer time and the Australian Research Council (Grants DP0984806 and DP1094170) for funding.
\end{acknowledgments}

\end{document}